\documentclass[12pt]{iopart}
\usepackage{comment}
\usepackage{inputenc}
\usepackage{graphicx}
\usepackage{bm}
\usepackage{mathenv}
\usepackage{nicefrac}
\usepackage{braket}
\usepackage[dvipsnames]{xcolor}

\begin{document}

\title[Rydberg ions in coherent motional states]{Rydberg ions in coherent motional states: A new method for determining the polarizability of Rydberg ions} 

\author{Marie Niederländer~$^{1}$, Jonas Vogel~$^{1}$, Alexander Schulze-Makuch~$^{1}$, Bastien Gély~$^{1}$, Arezoo Mokhberi~$^{1}$ and Ferdinand Schmidt-Kaler~$^{1,2}$}

\address{$^1$~QUANTUM, Johannes Gutenberg-Universität Mainz, 55128 Mainz, Germany}
\address{$^2$~Helmholtz Institut Mainz, 55128 Mainz, Germany}

\eads{\mailto{arezoo.mokhberi@uni-mainz.de}}

\vspace{10pt}
\begin{indented}
\item[\today]
\end{indented}

\begin{abstract}
We present a method for measuring the polarizability of Rydberg ions confined in the harmonic potential of a Paul trap. For a highly excited electronic state, the coupling between the electronic wave function and the trapping field modifies the excitation probability depending on the motional state of the ion. This interaction strongly depends on the polarizability of the excited state and manifests itself in the state-dependent secular frequencies of the ion. We initialize a single trapped $^{40}$Ca$^+$ ion from the motional ground state into coherent states with $|\alpha|$ up to 12 using electric voltages on the trap segments. The internal state, firstly initialised in the long-lived 3D$_{5/2}$ state, is excited to a Rydberg S$_{1/2}$-state via the 5P$_{3/2}$ state in a two-photon process. We probe the depletion of the 3D$_{5/2}$ state owing to the Rydberg excitation followed by a decay into the internal ground 4S$_{1/2}$ state. By analysing the obtained spectra we extract the polarizability of Rydberg states which agree with numerical calculations. The method is easy-to-implement and applicable to different Rydberg states regardless of their principal or angular quantum numbers. An accurate value of the state polarizability is needed for quantum gate operations with Rydberg ion crystals.
\end{abstract}

\maketitle

\section{Introduction}
\label{sec:introduction}

The electric dipole polarizability of atoms in highly excited electronic states, referred to as Rydberg states, scales with their principal quantum number to the power of seven~\cite{Gallagher1994}. This property of Rydberg atoms puts them forward as sensors for static and oscillating electric fields with unprecedented sensitivity~\cite{Haroche2006, Fan2015} and for the microwave field~\cite{Gordon2014, Anderson2016}. Along with significantly enhanced polarizability, long-range dipolar interactions between Rydberg atoms have opened up possibilities for exploring strongly interacting systems~\cite{Bloch2008}. More widely, applications of coherently controllable Rydberg atoms in quantum computing~\cite{Jaksch2000,Lukin2001,Saffman2010}, networking~\cite{Spong2021} and simulation~\cite{Weimer2010} have inspired many experiments with neutral atoms in optical lattices~\cite{Wilk2010, Gross2017, Bernien2017} and ultra-cold~\cite{Nipper2012} and room-temperature gases~\cite{Urvoy2015}. 

As an interesting system for exploiting Rydberg properties for such applications, trapped ions exhibit a number of advantages owing to the possibility for a superb control over their internal and external degrees of freedom~\cite{Mueller2008, Mokhberi2020, Feldker2015}. Additionally, advances for preparation, readout and entanglement operations have enabled successful demonstration of fault-tolerant protocols for quantum error correction~\cite{Hilder2021}.
To date, experimental work with Rydberg trapped ions is implemented in Paul traps, in which a combination of a static and an oscillating electric field generates the trapping potential~\cite{Leibfried2003}, and Rydberg states can be addressed using two ultra violet~(UV) lasers via an intermediate state in a two-photon process~\cite{Higgins2017, Andrijauskas2020}.

For an ion in the harmonic potential of a Paul trap, the Rydberg excitation energy is altered by the occupation of the modes of the ion motion because of the ion's large polarizability, which outperforms that of the ground state by, e.g., eight orders of magnitude in a specific case~\cite{Higgins2020}. This effect is explained by the quadratic Stark shift induced by the residual electric field at the ion position. In addition, this interaction leads to a state-dependent force on the ion in a Rydberg state that manifests itself in the stiffness of the trapping potential depending on the sign of the polarizability of the excited state~\cite{Mokhberi2020}. 
This effect has adverse consequences for implementing fast gates with Rydberg ions that use a dipole-induced force~\cite{Zhang2020,Li2013} as well as for atomic clocks and quantum logic experiments with two ion species, where a non-zero differential polarizability causes inaccuracy~\cite{Safronova2011, Brewer2019}. On the positive side, this effect can be used for sub-microsecond quantum gate operations with a two-ion crystal subjected to electric pulses in a Paul trap~\cite{Vogel2019}. This gate operation has potential to be extended to a multi-ion crystal and be further optimised by engineering the acquired phase and motional excitation using complex pulses to be operated much faster~\cite{HanBao2022}. 

Taking advantage of this effect, we developed a new method for determining the static dipole polarizability of Rydberg ions. We prepare a single trapped ion in a certain coherent motional state and subsequently probe the excitation spectrum for a Rydberg state. By analysing the transition frequency shift and the lineshape, we extract the polarizability of the excited Rydberg state. The method benefits from two techniques in Paul traps, each of which can be implemented with near-to-perfect fidelity~\cite{RoosPhD}:
first, reliably probing the internal state, and thus, the population transfer to a Rydberg state using fluorescence detection techniques in each experimental sequence, and second, reliable preparation of a certain motional state that allows for assigning the ion thermal distribution used in the analyses of Rydberg spectra for different motional states.
As an advantage, this method does not involve preparing Fock states of the ion motion, which is demanding as compared to preparing coherent states~\cite{Meekhof1996, Ziesel2013}. In our experiment, coherent states of the ion motion are generated using electric pulses. Such electric pulses reshape the trapping potential, displace the ion from the trap centre and result into the excitation of motional modes. 

This paper proceeds as follows. In section~\ref{sec:phonon-distribution}, we introduce our theoretical description for a single Rydberg ion in a Paul trap including its motional degree of freedom, which follows from our model for the spectrum of Rydberg ions in coherent motional states. In section~\ref{sec:experiment}, we present a generic ion trap and the relevant electronic setup that we used for measuring the polarizability of three different Rydberg S$_{\nicefrac{1}{2}}$ states of trapped $^{40}$Ca$^+$ ions. In section~\ref{sec:results}, experimental results and analyses based on the model described in section~\ref{sec:phonon-distribution} are discussed. Finally, a summary of main results and an outlook for applications of this method is given in section~\ref{sec:conclusion}.  

\section{Phonon distribution and Rydberg transitions}
\label{sec:phonon-distribution}

We consider a singly-charged ion in a Rydberg state in a standard linear Paul trap. The time-dependent trapping potential of the quadrupole electric field can be written as $\Phi(\bm{R},t) = \gamma_\mathrm{RF} (X^2-Y^2)\cos{(\Omega_\mathrm{RF} t)}-\gamma_\mathrm{DC}((1+\epsilon)X^2+(1-\epsilon)Y^2-2Z^2)$, with $\gamma_\mathrm{DC}$ and $\gamma_\mathrm{RF}$ being the gradients of static and oscillating potentials and $\Omega_\mathrm{RF}$ being the radio-frequency~(RF). A non-zero value of $\epsilon$ breaks the degeneracy of the two modes of the ion oscillation in the $x-y$ plane, and is determined from the trap geometry and applied voltages. The Hamiltonian of the system is given by \cite{Mueller2008}
\begin{eqnarray}
    H &=& H_\mathrm{CM} + H_\mathrm{el} + H_\mathrm{CM - el},
\end{eqnarray}
where
\begin{eqnarray}
    H_\mathrm{CM} &=& \frac{\bm{P}^2}{2M}+ \frac{1}{2}M \omega_x^2 X^2 + \frac{1}{2}M \omega_y^2 Y^2  + \frac{1}{2}M \omega_z^2 Z^2 , \\
    H_\mathrm{el} &=& \frac{\bm{p}^2}{2m}+ V(|\bm{r}|) + V_\mathrm{ls} - e\Phi(\bm{r},t)  , \\
    H_\mathrm{CM-el} &=& -2 e \gamma_\mathrm{RF} \cos{(\Omega_\mathrm{RF} t)}(Xx -Yy) -2 e \gamma_\mathrm{DC} \left((1+\epsilon)Xx + (1-\epsilon)Yy - 2Zz \right). \quad 
\end{eqnarray}
Here, $M$~($m$), $\bm{P}$~($\bm{p}$) and $\bm{R}=\{X,Y,Z\}$~($\bm{r}=\{x,y,z\}$) denote the mass, the momentum and the position of the ion core~(the position of the electron with a unit charge $e$). The term $H_\mathrm{CM}$ accounts for the ion core interaction in the trapping field as a three-dimensional harmonic oscillator with $\omega_i$ ($ i\in \{x, y, z\} $) frequencies. The term $H_\mathrm{el}$ denotes the Rydberg electron interactions, consisting of the electron kinetic energy, the Coulomb interaction with the doubly-charged ion core $V(|\bm{r}|)$, the spin-orbit coupling $V_\mathrm{ls}$ and its coupling to the trapping field $- e\Phi(\bm{r},t)$. Note that the intrinsic micro-motion of the ion is negligible here since the RF frequency is typically about ten times larger than secular frequencies \cite{Mueller2008, Cook1985}.
The term $H_\mathrm{CM-el}$ accounts for the coupling of the electronic dynamics to the ion core because of the trapping fields. This interaction is the quadratic Stark effect that manifests itself as a state-dependent force on the ion and couples its internal and external degrees of freedom. Since the electron dynamics is much faster than the ion motion, this term can be treated using second-order perturbation theory~\cite{Mueller2008}. The modified secular frequencies $\omega^{\prime} _i$ can be written as~\cite{Mokhberi2020}
\begin{eqnarray}
\label{eq:state-dep-freq1}
    \omega_{x,y}' &=&\sqrt{\frac{2 e^2 \gamma_\mathrm{RF}^2}{M^2 \Omega_\mathrm{RF}^2}-\frac{2 e \gamma_\mathrm{DC} (1 \pm \epsilon)}{M}-\frac{2 \mathcal{P} \left(\gamma_\mathrm{RF}^2 +\gamma_\mathrm{DC}^2 (1 \pm \epsilon)^2\right)}{M}},
\end{eqnarray}
\begin{eqnarray}
\label{eq:state-dep-freq2}
    \omega_z' &=& \sqrt{\frac{4 e   \gamma_\mathrm{DC}}{M} - \frac{16 \mathcal{P} \gamma_\mathrm{DC}^2}{M}}.
\end{eqnarray}
The state polarizability $\mathcal{P}$ scales with the principal quantum number to the power of seven, and for a Rydberg state, can lead to a considerable modification of the trapping frequencies~$\omega_{i}'$, e.g., in the order of a few percent as measured for 50 to 66 S-states of $^{40}$Ca$^+$~\cite{Andrijauskas2020}. The sign of this shift depends on the sign of $\mathcal{P}$, and Rydberg dressed states with opposite polarizabilities can be used to cancel this shift~\cite{Li2013}.

The aim of this work is to extract the polarizability of Rydberg ions using the manifestation of this Stark effect in the transition spectra for a trapped ion in different motional states. The Rydberg lineshape for an ion in an excited state of motion can be written as a convolution of all lineshapes for each occupied mode weighted by the phonon distribution $P_n$ for that specific motional mode: 
\begin{equation}
\mathcal{L}^{\prime} (\omega) = \prod\limits_\mathrm{n=0} P_n \mathcal{L}(\omega + n\Delta \omega, \sigma, \Gamma).
\label{eq:transitionLineshape}
\end{equation}
Here, $\mathcal{L}$ denotes the lineshape as a {\it Voigt} profile that accounts for the modification of trapping frequencies~$\Delta \omega= \omega_{i}'-\omega_{i}$, Doppler broadening~$\sigma$ due to the laser(s) addressing the transition and the linewidth(s) $\Gamma$ of the corresponding laser(s) used for the single- or multi-photon Rydberg excitation. Note that the occupied modes cause an alternation of the line function in addition to the centre frequency shift.

We now proceed our discussion for coherent states of the ion motion in accordance with our experiments; however, the presented formulation can be adopted for any other motional state by inserting the relevant statistical distribution of the phonon number.  
A coherent state $\ket{\alpha}$ can be generated by displacing the ground state of a harmonic oscillator and can be written as an expansion in Fock space:
\begin{equation}
\ket{\alpha} = \exp \left( -\frac{1}{2}\left|\alpha\right|^2\right) \sum_{n=0}^\infty {\frac{\alpha^n}{\sqrt{n!}}\ket{n}},
\label{eq:coherentState}
\end{equation}
with the probability distribution for $n$ phonons given by
\begin{equation} \label{eq:coherentDistribution}
P_n(\alpha) = \frac{|\alpha|^{2n}}{n!}e^{-|\alpha|^2},
\end{equation}
with $|\alpha|$ being the coherent state size and $n$ the phonon number in the $\ket{n}$ state.
To generate coherent states, we consider the use of electric pulses in the form of many sinusoidal waves, which kick the ion out of its equilibrium position. By the action of switching voltages on trap electrodes, the trapping potential is successively reshaped such that the trap centre is shifted along the applied force for a kick pulse time and coherent states are generated.

To characterise the motional state of the ion in the laboratory, first the ion is prepared close to the ground state of its motion using sideband cooling techniques~\cite{Leibfried2003}. Coupling of the trapped-ion oscillator to a dipole forbidden transition by laser radiation allows for probing the motional state using resonant transitions with the Rabi frequency $\Omega_{n,n+s}$ for the $s^{\rm{th}}$ sideband. Note that the optical field of a probe laser couples the internal ground state of the atom to an upper state that is close to the ground state, which is a meta-stable D state in our experiment, as opposed to a Rydberg state (see section~\ref{sec:experiment}). By repeating this measurement sequence many times per laser probe time $\tau$, the probability to find the system in the upper state is given by  
\begin{equation}
\label{eq:ProbabilityD}
P_{\rm{D},s}(\tau)=\sum_{n=0} P_n \cdot \frac{1}{2} A \left[1-\cos(\Omega_{n,n+s} \tau )\right] \mathrm{e}^{-\gamma \tau},
\end{equation}
where $P_n$ is the number state distribution depending on the motional state of the ion in the corresponding motional mode and $\gamma$ is a universal decoherence rate for the given motional state. Here, we introduced $A$ as a free parameter in analysing experimental data that accounts for an offset to the base line of spectra and contributes to the strength of lines (see section~\ref{sec:results}). Note that $A$ scales with the maximum probability for a $\pi$-pulse of the optical field.

\begin{figure}[ht]
    \centering
    \includegraphics[scale=0.2]{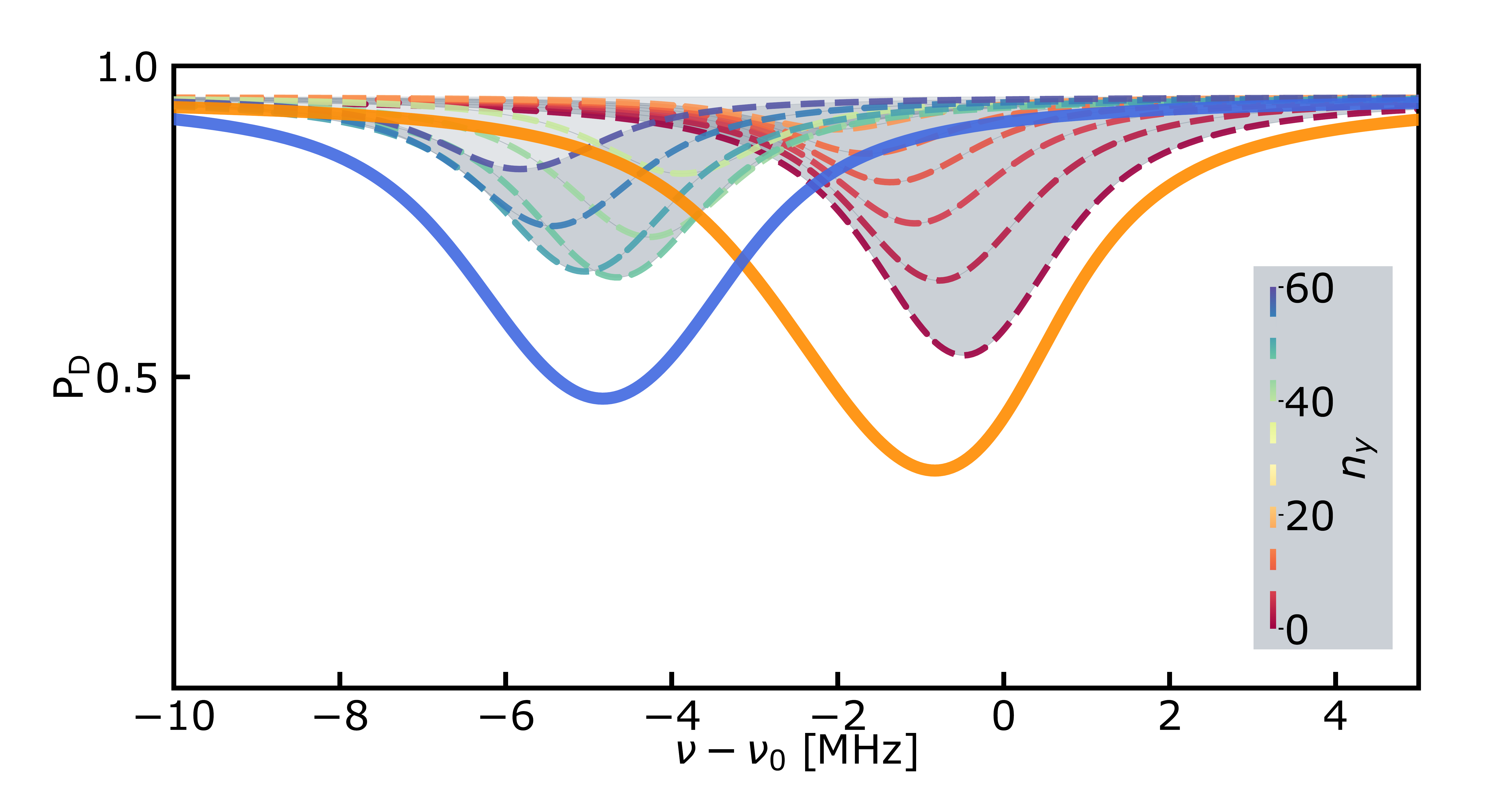}
    \caption{Simulation of the resonance spectra for the 57S$_{1/2}$ state of a trapped $^{40}$Ca$^+$ ion including its motional degree of freedom. The spectra for a thermal state  with $\bar{n}_x=\bar{n}_y=6$ (solid orange) and for one radial mode in a coherent state with $\bar{n}_y=|\alpha|^2=49$ and the other two modes in a thermal state with $\bar{n}_x=1$ (solid blue). These two spectra were calculated as a weighted sum over single Fock spectra (dashed profiles). Each Fock spectrum was weighted with phonon number probabilities in the two radial modes $x$ and $y$ and are a scale factor of 1.5 for better visibility. 
    For simplicity, only 6 Fock spectra taking part in convolution of the thermal and coherent spectra are shown, where phonon number in the $y$ mode, $n_y$ is indicated by the colour bar.}    \label{fig:lineshapemodel}
\end{figure}

A successful Rydberg excitation is detected by means of fluorescence detection techniques described in~\cite{Mokhberi2020} as the excited electronic state decays to the ground or near-ground states. Rydberg spectra simulated using our model for thermal and coherent states of a trapped $^{40}$Ca$^+$ ion are shown in Fig.~\ref{fig:lineshapemodel}.
For these calculations, we considered the two following cases to illustrate the effect of the phonon distribution when only one radial mode of the ion motion is differently occupied. First, a thermal distribution for three motional degrees of freedom with the mean phonon number $\bar{n}_x=\bar{n}_y=6$, which corresponds to a temperature of about $0.6$~mK, and second, a case of a thermal distribution for the $x$ mode at $\bar{n}_x=1$, corresponding to a temperature of about $0.1~$mK, while the $y$ mode is in a coherent state. The width of a single Fock spectrum is determined by the laser line width of about $2~$MHz (FWHM) and the temperature in the axial mode $z$, corresponding to $0.5~$mK.

\section{Experimental implementation}
\label{sec:experiment}

\subsection{Experimental setup}
\label{subsec:setup}

In our experiments, we use $^{40}$Ca$^+$ ions confined in a linear Paul trap that features an optical access through pierced endcap electrodes for the lasers addressing Rydberg transitions, and in addition, segmented DC electrodes used for generating electric pulses. The former is important for minimising Doppler broadening of Rydberg lines using two counter-propagating laser beams that address the transition and the latter is needed for generating coherent states \cite{Alonso2016}.
Neutral calcium was isotope-selectively photoionized using $375$~nm and $423$~nm~\cite{Wolf2018} laser light in a zone for loading ions. Ions were moved to an experimental zone, $1.5$~mm away from the trap centre, to avoid contamination of central electrodes. Secular trapping frequencies in the experimental zone were set at $\{ \omega_x,\omega_y,\omega_z \}=2\pi \times \{2.16,1.8,1.05\}$~MHz, where a radio-frequency~(RF) of $2\pi \times 14.11$~MHz was applied to RF electrodes. A set of four electrodes, which are elongated along the trap axis $z$, was used to precisely position ions in the plane perpendicular to the trap axis, and hence to minimise residual stray electric fields. Static voltages for transporting ions were provided by an arbitrary waveform generator~(AWG). To excite motional states of the ion, we used another AWG for mixing an external radio frequency voltage with DC voltages applied to one of the elongated electrodes, see Fig.~\ref{fig:electric_setup}.

\begin{figure}
    \centering
    \includegraphics[scale=0.23]{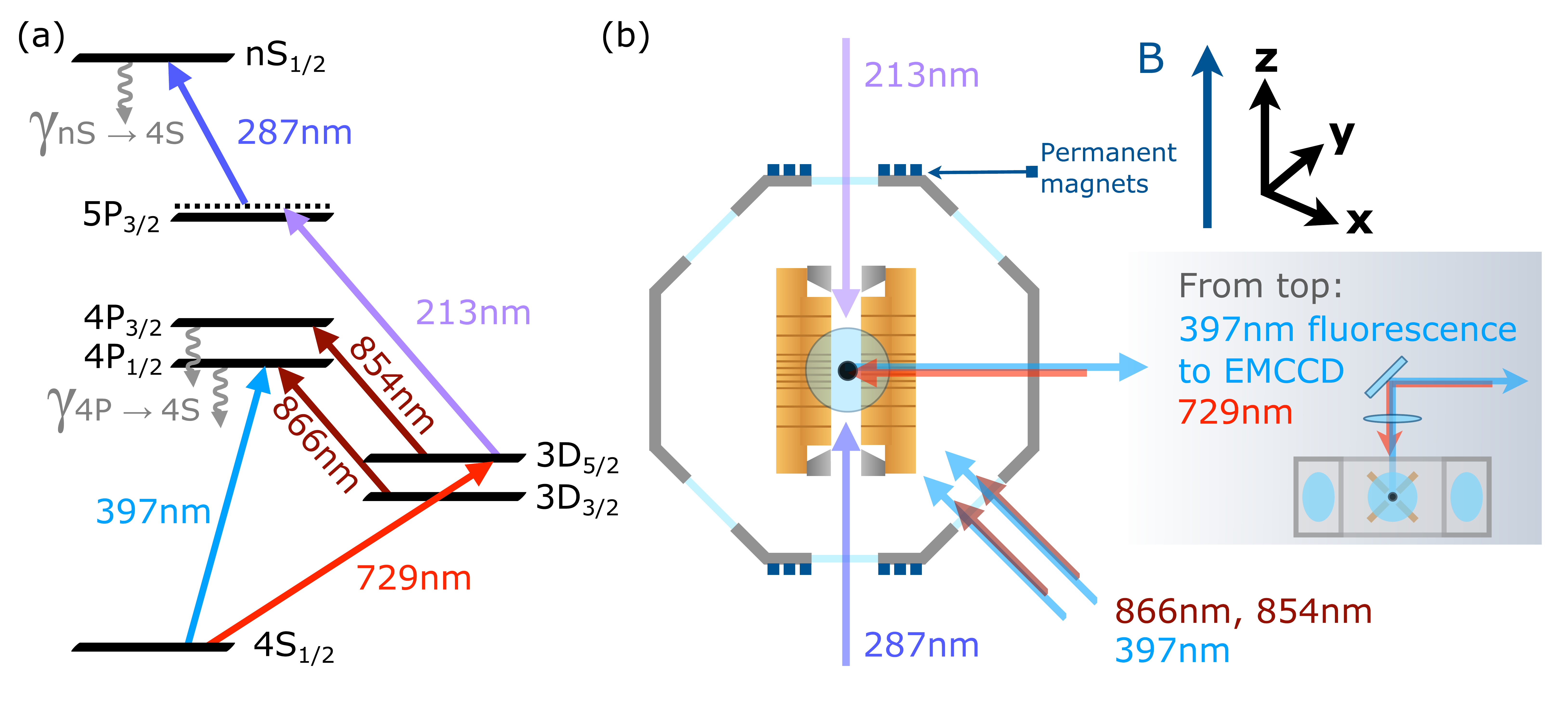}
    \caption{(a) Reduced energy level diagram of $^{40}$Ca$^+$ with relevant wavelengths for laser cooling, state initialization, Rydberg excitation and state detection. Decays from a Rydberg S$_{\nicefrac{1}{2}}$ state to the 4S$_{\nicefrac{1}{2}}$ state as well as from the 4P$_{\nicefrac{1}{2}}$ and 4P$_{\nicefrac{3}{2}}$ states are shown with grey wavy lines. Permanent magnets (in dark blue) were used for generating a 0.55 mT magnetic field along the trap axis. Zeeman sublevels of the 4S$_{\nicefrac{1}{2}}\rightarrow$3D$_{\nicefrac{5}{2}}$ transition were used for sideband cooling and for determining the size of coherent motional states. (b) Ion trap in the center of a hexagonal vacuum chamber with optical access for laser beams and fluorescence detection using an EMCCD camera. For Rydberg excitation, laser beams at $213$~nm and $287$~nm, which are counter-propagating along the trap axis, pass through endcap electrode holes.}
    \label{fig:setup}
\end{figure}

Doppler cooling was implemented using the 4S$_{\nicefrac{1}{2}}\rightarrow$ 4P$_{\nicefrac{1}{2}}$ transition by $397~$nm laser radiation. Laser sources near $854$ and $866~$nm depopulate the metastable 3D$_{\nicefrac{5}{2}}$ and 3D$_{\nicefrac{3}{2}}$ states, respectively. State initialization and sideband cooling is performed by coherent state transfer for the quadrupole 4S$_{\nicefrac{1}{2}}\rightarrow$ 3D$_{\nicefrac{5}{2}}$ transition with $729~$nm laser, which was stabilized to a high finesse cavity. Relevant energy levels and laser transitions are shown in Fig.~\ref{fig:setup}(a).
Sideband cooling of the radial motional modes was carried out using the Zeeman sublevels of the 4S$_{\nicefrac{1}{2}}(m_J=\nicefrac{-1}{2})\rightarrow$ 3D$_{\nicefrac{5}{2}}(m_J=\nicefrac{-5}{2})$ transition using a $729$~nm laser beam, which results into an average thermal phonon number $<0.5$.

For initialization in the Zeeman sublevel of the electronic ground state with $m=\nicefrac{-1}{2}$, we used eight $\pi-$pulses on the 4S$_{\nicefrac{1}{2}}(m_J=\nicefrac{1}{2})\rightarrow$ 3D$_{\nicefrac{5}{2}}(m_J=\nicefrac{-3}{2})$ transition with consecutive quench using laser light near $854~$nm. The ion is initialized in the 3D$_{\nicefrac{5}{2}}(m_J=\nicefrac{-5}{2})$ state by rapid adiabatic passage (RAP) on the 4S$_{\nicefrac{1}{2}}(m_J=\nicefrac{-1}{2})\rightarrow$ 3D$_{\nicefrac{5}{2}}(m_J=\nicefrac{-5}{2})$ transition. 

\begin{figure}
    \centering
    \includegraphics[scale=0.22]{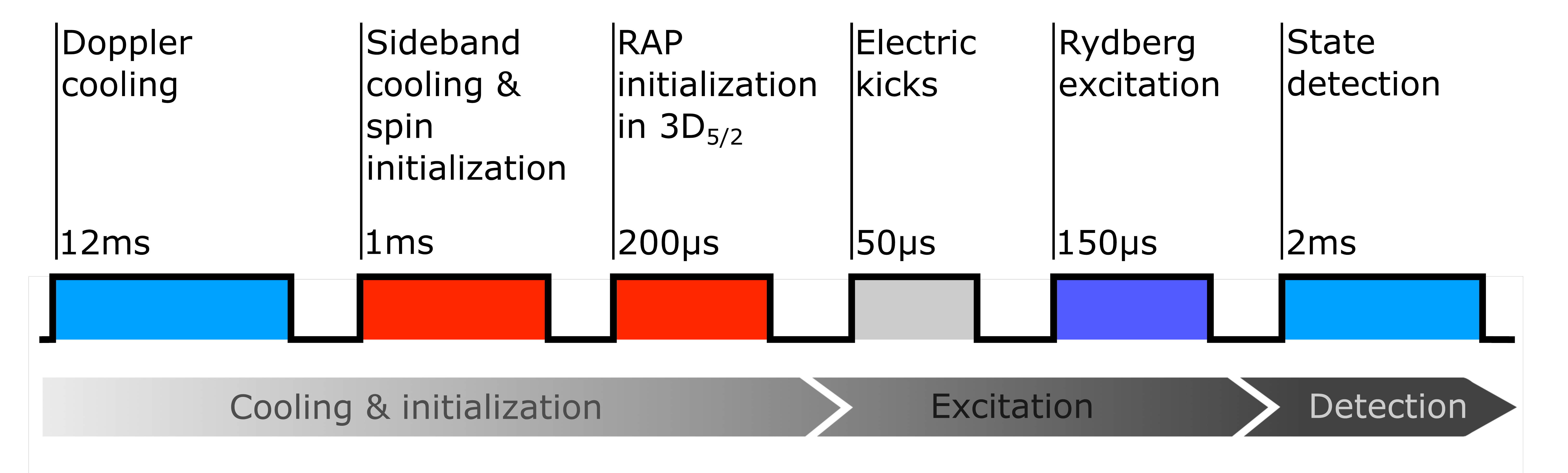}
    \caption{Pulse scheme for the experimental sequence labelled with the time duration applied in the experiment. Cooling and state initialization were carried out using the $397$~nm (blue) and $729$~nm lasers (red). Coherent motional states were generated by applying an electric kick (in grey). Rydberg states were excited in a resonant two-photon process using $213$~nm and $287$~nm lasers (violet). The final state is detected using fluorescence detection techniques (see also Fig.~\ref{fig:setup}).}
    \label{fig:sequence}
\end{figure}

State detection is performed by collecting scattered fluorescence near $397~$nm using an complex objective lens with N.A.$=0.28$ and imaging on an electron-multiplying charge coupled device (EMCCD) camera. The ion does not scatter light in the 3D$_{\nicefrac{5}{2}}$ state, but in the 4S$_{\nicefrac{1}{2}}$ state scatters light that is an indication of a successful Rydberg excitation. A schematic of the experimental sequence including the electric kick to excite coherent states of motion and typical timings is shown in Fig.~\ref{fig:sequence}, which is typically repeated 100 times.

\subsection{Excitation of coherent states}
\label{subsec:motional-excitation}

\begin{figure}
    \centering
    \includegraphics[scale=0.23]{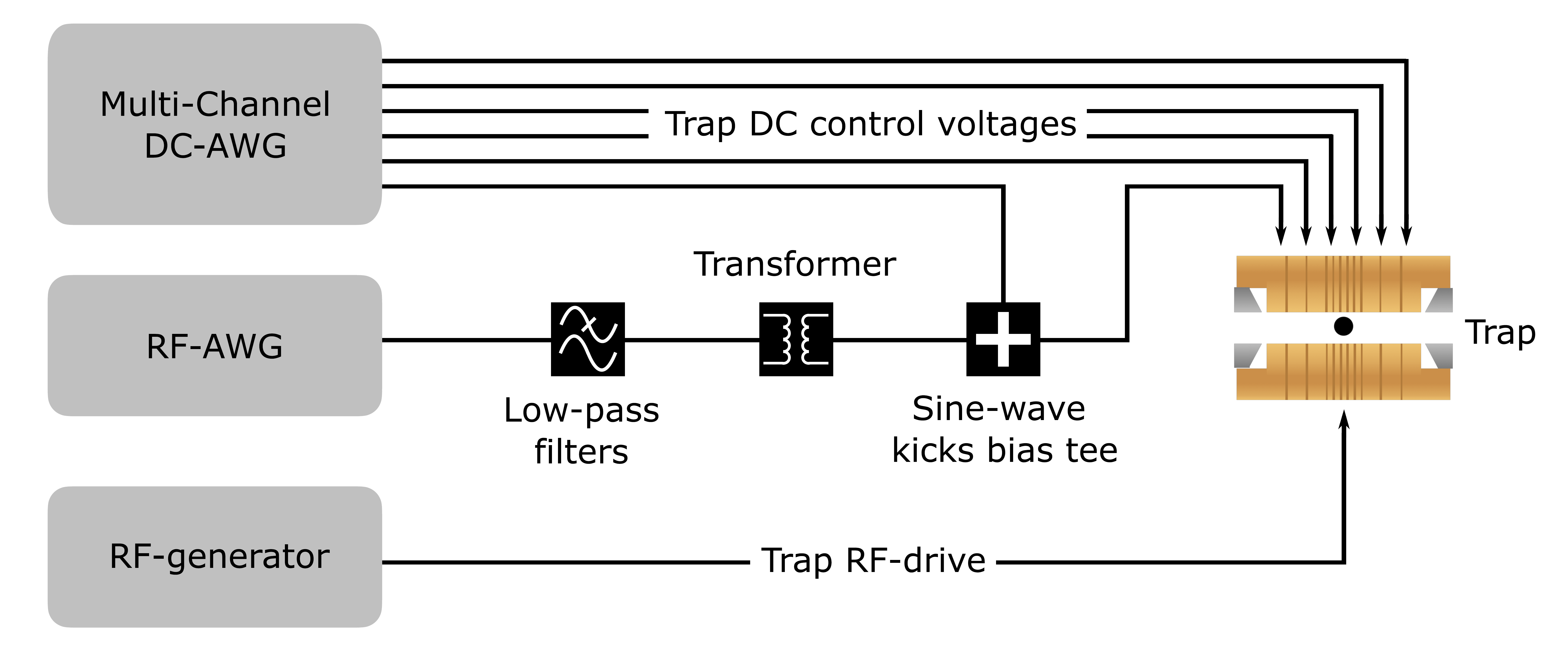}
    \caption{Schematic of the electronic setup. For applying fast electric pulses on trapped ions, we used two arbitrary waveform generators (AWGs), low-pass filters for each DC electrodes and a 1:1 ratio transformer for isolating the RF-AWG from noise. Simultaneous application of DC voltages and sinusoidal pulses on one trap electrode elongated along the trap axis was carried out by feeding them through a bias tee.}
    \label{fig:electric_setup}
\end{figure}

\begin{figure}
    \centering
    \includegraphics[scale=0.23]{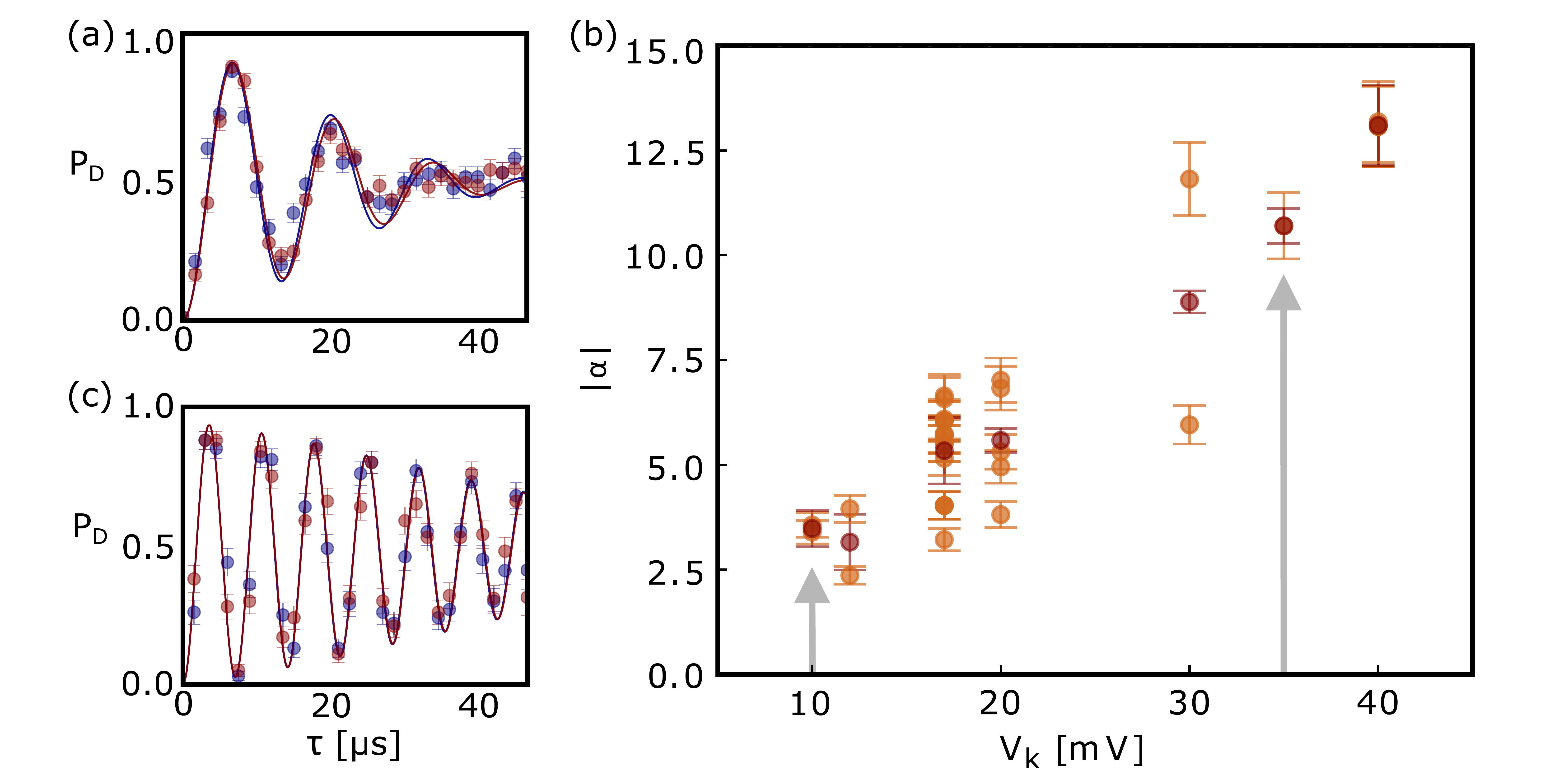}
    \caption{Calibration of the electric kick amplitude for determining coherent state sizes. (a,c) Transition probabilities for the first red (red dots) and the first blue (blue dots) sidebands of the 4S$_{\nicefrac{1}{2}}(m_J=\nicefrac{-1}{2})\rightarrow $3D$_{\nicefrac{5}{2}}(m_J=\nicefrac{-5}{2})$ transition as a function of the $729$~nm-laser pulse width when the electric pulse amplitude was $V_k=10$~mV and $35$~mV, respectively. Fits are used for extracting  $|\alpha|$ (solid lines). (b) Resulting values for $|\alpha|$ as a function of $V_k$ measured for 100 sine pulses (light brown) and mean values (dark brown). Error bars depict a one-$\sigma$ uncertainty interval for $|\alpha|$ obtained from a Monte-Carlo simulation. The spread in $|\alpha|$ results from the frequency selectivity of the electric sine-pulse in combination with kHz-drifts of the trap frequencies.}
    \label{fig:KickCalibration}
\end{figure}

To generate coherent states of the radial motional mode along the $y$ direction, we used an AWG (Hameg HMF2550) to drive a hundred-sine-wave pulse with the frequency $\omega_{y}$. The amplitude of the kicking pulse $V_{k}$ can be varied in the range of $10-40$~mV, allowing for fine tuning of the coherent state size between $\left|\alpha\right|\in [2,11]$. The kicking pulse was applied to one elongated electrode along the trap axis. As a result, coherent motional states are generated only in one radial mode, the $y$ direction, while the other two modes remained in their thermal states. Fig.~\ref{fig:electric_setup} shows the schematic of our electronic setup.
As discussed in section~\ref{sec:phonon-distribution}, this electric pulse leads to the excitation of coherent motional modes with an amplitude that is a function of the electric kick amplitude and its duration. 
We calibrate the strength of the coherent excitation using sideband spectroscopy on the 4S$_{\nicefrac{1}{2}}(m_J=\nicefrac{-1}{2})\rightarrow $3D$_{\nicefrac{5}{2}}(m_J=\nicefrac{-5}{2})$ transition as a function of the laser pulse amplitude and pulse duration addressing this transition (Fig.~\ref{fig:KickCalibration}). By varying the duration of the electric pulse we are able to precisely measure the ion secular frequency along the kick direction. Note that for integer multiples of the ion's secular frequency the motional excitation after the electric field kick gets back to zero again. Residual phonons are due to a non-zero heating rate of about  $<0.2~$phonons/ms for the radial mode along the $y$ axis and due to a non-zero angle that might occur between the kick direction and the principal axis of the trap.

To determine the coherent state size $|\alpha|$, we first extract the thermal phonon numbers and the coupling strength from a fit to the carrier and red and blue sideband Rabi oscillations for the thermal ion on the 4S$_{\nicefrac{1}{2}}(m_J=\nicefrac{-1}{2})\rightarrow $3D$_{\nicefrac{5}{2}}(m_J=\nicefrac{-5}{2})$ transition \cite{Leibfried2003}. Figs.~\ref{fig:KickCalibration}(a,c) show red and blue sideband pulse-width-scans measured for $V_{k}=10~$and $35$~mV and the corresponding fits.
The coherent state size was obtained from a common fit to the red and blue sideband after the electric kick, assuming $P_n$ to be the coherent distribution (Eqn.~\ref{eq:coherentDistribution}) and $\Omega_0$ given by the thermal fit.  A large variance in $|\alpha|$ for the same $V_{k}$ is due to the frequency selectivity of the electric kick and drifts of the trap frequency due to a drift of the RF-drive amplitude in the order of $1-2 \%$. As a consequence, a few kHz of frequency drifts may cause a drastically large uncertainty for determining the coherent state size. For this reason, one calibration measurement of $|\alpha|$ was not sufficient for all kick voltages, and therefore, $|\alpha|$ was determined separately for each coherent excitation in each measurement.

To calculate the uncertainty in $|\alpha|$, we simulated noisy data sets for a Rabi frequency of the carrier transition at the motional ground state and the Lamb-Dicke parameter $\eta$, in which values corresponding to our experimental settings were used. One data set contains a set of carrier, red and blue sideband Rabi oscillations for a thermal ion and a set of red and blue sideband Rabi oscillations for an ion in a coherent state of size $|\alpha|$.
The dominant error arises from the uncertainty in determining $\eta$, since an accurate measurement of the angle between the trap axis and the laser beam that addresses the quadrupole transition at $729~$nm was not available. For the angle of $\pi/4$ arranged in our setup, we considered an uncertainty of about $10\%$. 
The values obtained from the simulation of a thermal and a coherent motional state were then fitted to the same model that was used for analysing the experimental data. 
Start values for the fits were chosen from normal distributions of $\eta$ around $\pi/4$ with $\sigma_{\eta}=\pi/40$. The start values for $|\alpha|^2$ were taken from a normal distribution with $\sigma=\sqrt{|\alpha|}$. 
The fits were computed 2000 to 3000 times for each data set and the simulation was repeated for different values of $|\alpha|\in \left[3,10\right]$. The result is a relative uncertainty of $\sigma_{\mathrm{|\alpha|}}=7.7(5)\%$ (Fig.~\ref{fig:KickCalibration}(c)).

\subsection{Rydberg excitation of ions in coherent motional states}
\label{subsec:rydberg-excitation}

Excitation to Rydberg S$_{\nicefrac{1}{2}}$ states with principal quantum numbers between 35 and 70 was implemented in a resonant two-photon process using $213$ and $287~$nm lasers~\cite{Andrijauskas2020}. These Rydberg states mainly decay into the electronic ground state within about $10$~$\mu$s that has been computed as the Rydberg state lifetime~\cite{Glukhov2013}. The Rydberg excitation was probed by the population transfer to the metastable 3D$_{\nicefrac{5}{2}}$ state and if this state is not depopulated by the Rydberg excitation, the ion does not scatter light during the detection time of typically $2~$ms.

After preparation of the ion in a certain coherent state, we excited the ion into 49, 53 and 57 S$_{\nicefrac{1}{2}}$ states from the 3D$_{\nicefrac{5}{2}} (m_J=\nicefrac{-5}{2})$ state and via the intermediate 5P$_{\nicefrac{3}{2}}$ state in a resonant two-photon process.
Excitation of the intermediate state has to be avoided to prevent any decay due to the short lifetime of 5P$_{\nicefrac{3}{2}}$ state, which is about $35$~ns \cite{Safronova2011}. Therefore, the $213$~nm laser was detuned from the 5P$_{\nicefrac{3}{2}}$ state by about  $\Delta_{213} = 30 - 60 ~$MHz.

\section{Results}
\label{sec:results}
\subsection{Characterization of Rydberg spectra including motional states}
\label{subsec:characterization}

Depopulation of the 3D$_{\nicefrac{5}{2}}$ state as a function of the two-photon detuning gives the spectrum of a Rydberg state as described in Eqn.~\ref{eq:transitionLineshape}. Phonon distributions for both radial motional modes, the residual Doppler broadening along the trap axis and the finite laser linewidth affect the lineshape. The natural line width, excitation laser widths and a residual Doppler broadening due to a momentum transfer were taken into account in a Voigt profile~\cite{Andrijauskas2021}.
Engineering the phonon distribution prior to Rydberg excitation reduces the free fit parameters to the center frequency and the polarizability. To minimise residual frequency fluctuations due to non-zero laser width for Rydberg excitation, we evaluated relative shifts between different phonon distributions. Here, Rydberg excitation of an ion with radial phonon distributions close to the motional ground state were probed and convoluted with motional distributions as described above. From a correlated fit of both transition lineshapes the center frequency and the polarizability were extracted. 


\begin{figure}
    \centering
    \includegraphics[scale=0.23]{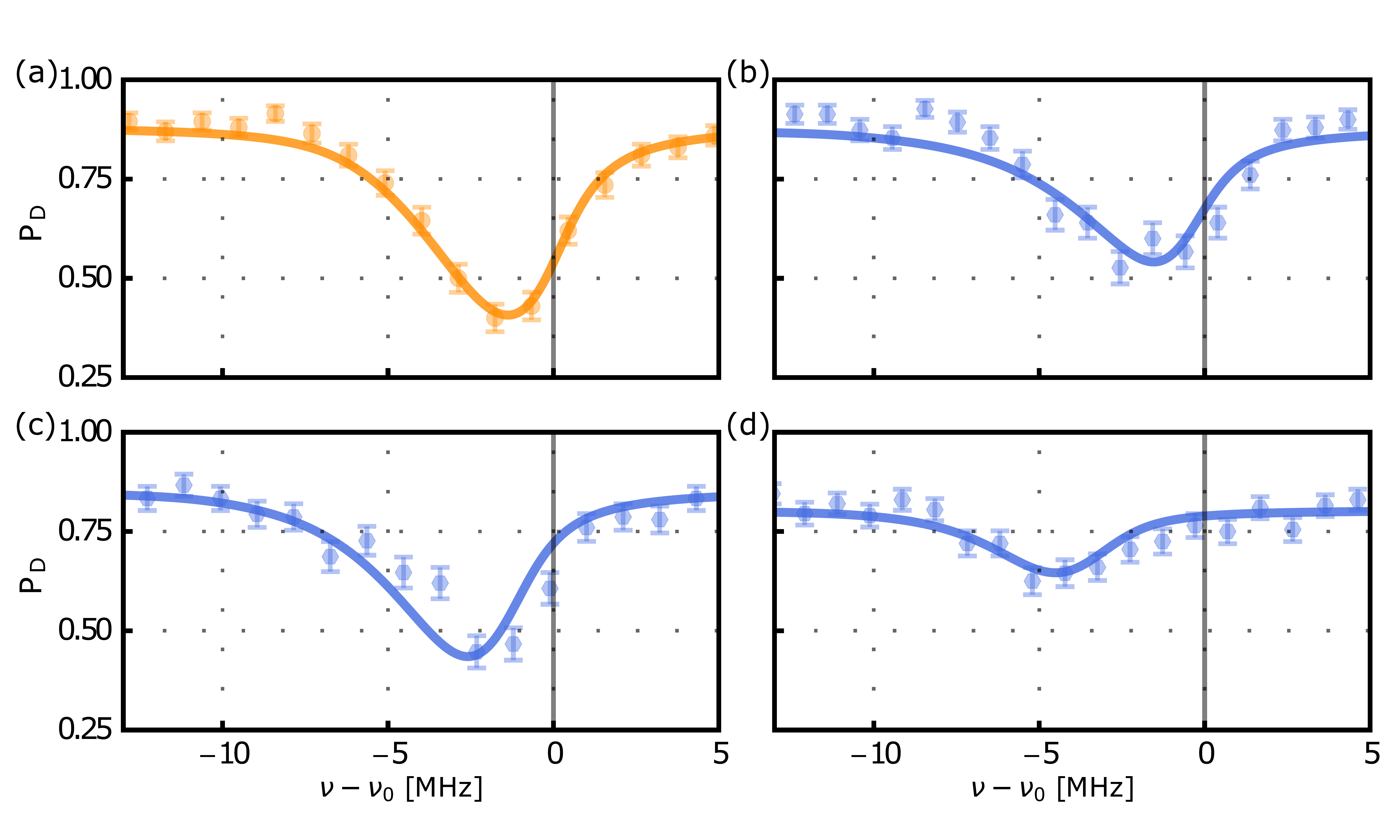}
	\caption{Depopulation spectra for the 3D$_{\nicefrac{5}{2}}(m_J=-5/2) \rightarrow 57$S$_{\nicefrac{1}{2}}(m_J=-1/2)$ transition of: (a) the ion in a thermal state and (b,c,d) the ion excited to coherent motional states of only one radial mode ($y$) with $\left|\alpha\right|=\{2.4(2),3.8(3),6.0(5)\}$ while the other two modes are in a thermal state. Error bars result from statistical noise, and the fit to the model function corresponds Eqn.~\ref{eq:transitionLineshape} (solid line). The shift of the excitation energy is induced by the state-dependent trap frequencies. The corresponding shifts for (a-d) are $\nu-\nu_0 = \{-1.74(1),-2.1(2),-2.8(2),-4.3(3)\}$ which agree with calculated values from Eqns.~(\ref{eq:state-dep-freq1}) and~(\ref{eq:state-dep-freq2}).
    \label{fig:PolarizabilityFit}}
\end{figure}

Depopulation spectra for the 57S$_{\nicefrac{1}{2}}$ Rydberg state as a function of two-photon detuning are shown in Fig.~\ref{fig:PolarizabilityFit}(a-d).
For a purely thermal state close to the ground state, residual phonons in both radial modes lead to a frequency shift of $\nu-\nu_0 = 1.74(1)~$MHz as compared to the two-photon resonance. For a thermal state and three coherent states with $\left|\alpha\right|=\{2.4(2),3.8(3),6.0(5)\}$, we observed frequency shifts of $\nu-\nu_0 = \{-1.74(1),-2.1(2),-2.8(2),-4.3(3)\}$, which increase with the coherent state size. These shifts are resulted from the state-dependent shift of trap frequencies for Rydberg states and correspond to a transfer of the ion's kinetic energy into the electronic state energy. 

\begin{figure}
    \centering
    \includegraphics[scale=0.23]{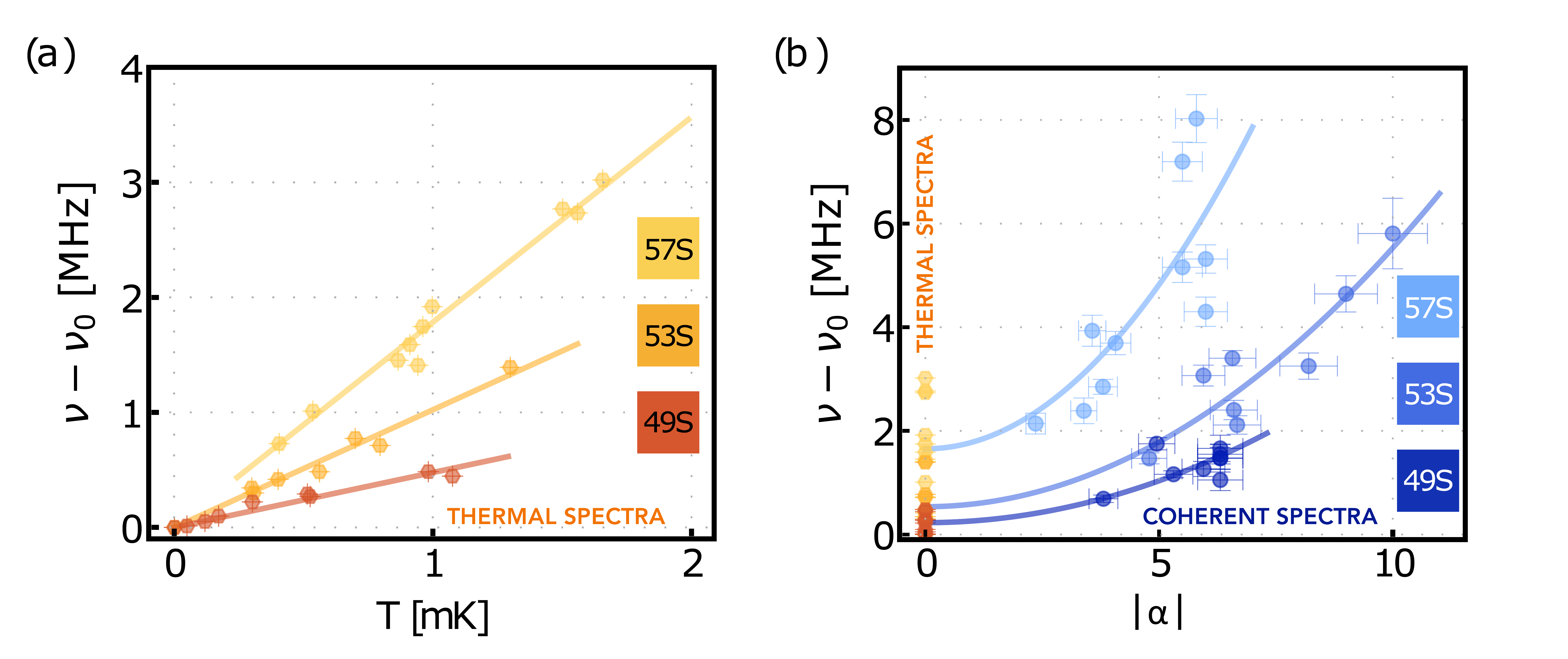}
	\caption{Absolute line shifts calculated from the model fit to measured spectra for the three Rydberg states. (a) For thermal motional states, the line shift exhibits a linear dependency to the average phonon number for both radial modes (solid lines show linear fits to data). Colour code- red, orange and yellow: 49S, 53S, and 57S states. (b) For coherent motional state, $|\alpha|$ dominates the line shift due to a much larger average phonon number in the excited mode along the $y$ axis as compared to the mode along the $x$ axis. The line shift has a quadratic dependency on $|\alpha|$ for different coherent states of the $y$ axis (solid curves show quadratic fits to data). Colour code- light, middle and dark blue: 49, 53, and 57S$_{\nicefrac{1}{2}}$ states. Purely thermal data is included as a reference for $|\alpha|=0$ in (a).  \label{fig:lineshift_results}}
\end{figure}

\subsection{Calculations of state polarizability}
\label{subsec:calculations}
\begin{figure}
    \centering
    \includegraphics[scale=0.23]{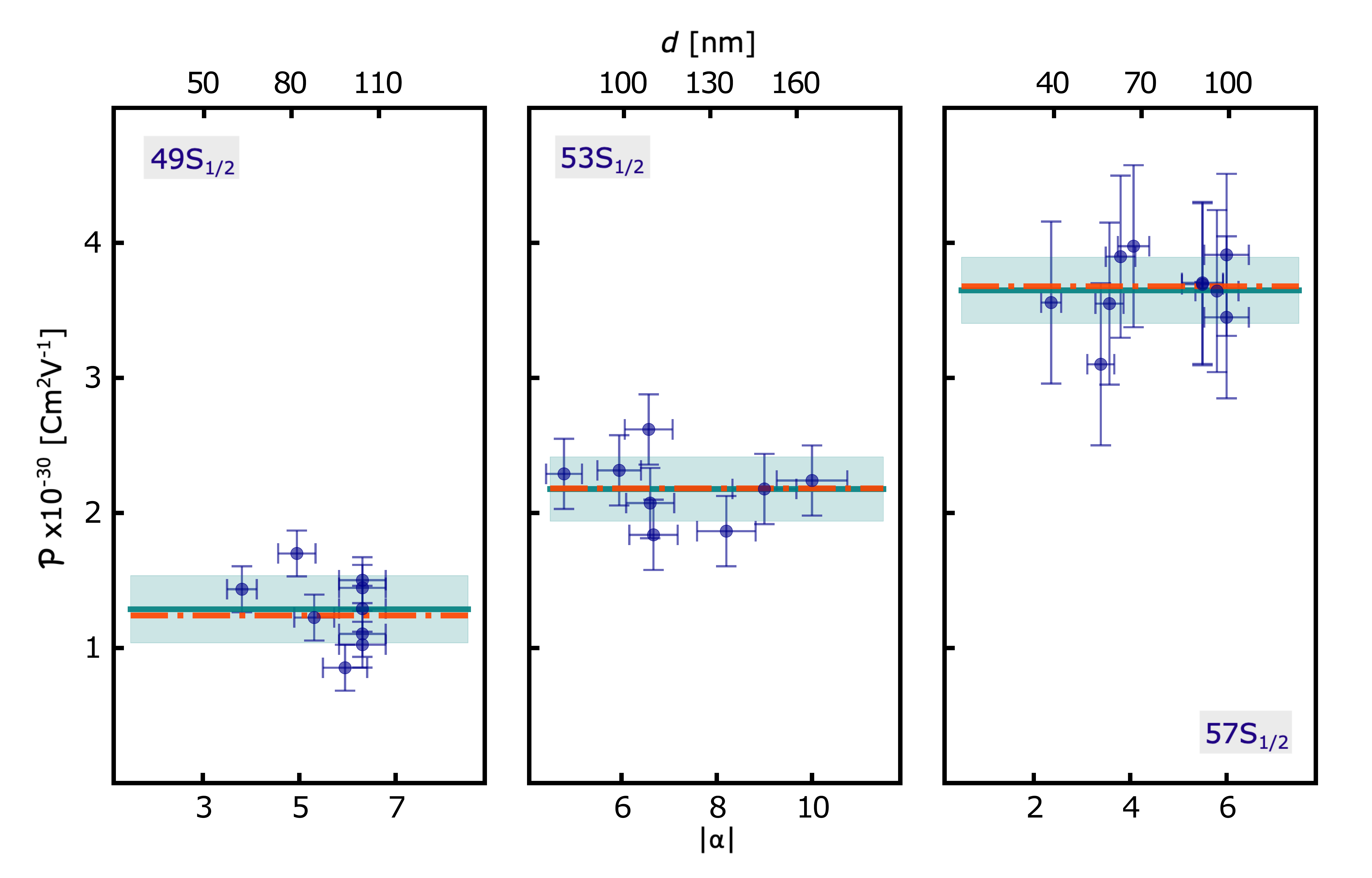}
	\caption{Polarizabilities of Rydberg 49S$_{\nicefrac{1}{2}}$, 53S$_{\nicefrac{1}{2}}$ and 57S$_{\nicefrac{1}{2}}$ states of $^{40}$Ca$^+$ as a function of the coherent state size $|\alpha|$ and as a function of the ion displacement from the trap centre. Error bars in y-axis are resulting from the accuracy of the line fits and in x-axis from the calibration of the kick amplitude. The data points for each state (blue) and mean value are shown within a 1$\sigma$ confidence band (green), together with the theoretical values from~\cite{Kamenski2014} (orange).}
    \label{fig:pol_vs_alpha}
\end{figure}

The lineshape model accounts for the underlying phonon distributions and a correlated fit of data allows to extract the ion's polarizability and the two-photon resonance frequency. Parameters as trap frequencies for the electronic ground state and the phonon numbers for thermal as well as coherent phonon distributions were determined through independent measurements.   
The polarizability of Rydberg S$_{\nicefrac{1}{2}}$ states with the principal quantum number 49, 53, and 57 were extracted using the Rydberg spectra obtained for various $\left| \alpha \right|$. Since the ion explores the electric fields of the trap, it is insightful to express the coherent state size in terms of the wave-packet extension of a harmonic oscillator. A thermal ion close to its motional ground state has an extension of a few nm and can be well approximated as stationary in the harmonic potential. Even for moderate coherent state sizes below $\left| \alpha \right|=10$ the extension of the wave packet exceeds $100~$nm in an electric field of about $\rm{\left| E \right|}=160$~V/m for a given set of trapping parameters. A Rydberg ion is more susceptible because of its higher polarizablity and additionally because of motional excitation, its wave packet has larger overlaps with the trapping fields. 

In calculations for the polarizability of each state, we mitigated systematic errors in our lineshape model by calculating the lineshape modifications for various coherent motional states. Results of these calculations are shown in Fig.~\ref{fig:pol_vs_alpha}. All individual measurements of the polarizability are weighted by the respective fit error and averaged for the three excited states. The resulting mean value is presented as green lines with shaded $1\sigma$ confidence bands and compared to literature values obtained by the second order Stark effect in Kratzer-Fues model potential calculations (orange) \cite{Kamenski2014}. Numerical values are given in table \ref{tab:results}. 
\begin{table}[ht]
    \begin{center}
        \begin{tabular}{|c|c|c|}
        \hline
          \multicolumn{3}{|c|}{$\mathcal{P}(\mathrm{\it{n}_{\rm{Ryd}}\rm{S}}_{\nicefrac{1}{2}})\times 10^{-30}~[\frac{\mathrm{Cm}^2}{\mathrm{V}}]$}\\
            \hline
           ~~$\it{n}_{\rm{Ryd}}$~~&~~This work~~&~~Theory~\cite{Kamenski2014}~~\\
            \hline
            $49$ &  1.3(2) & 1.24 \\ 
            \hline
            $53$ & 2.2(2) & 2.18 \\
            \hline
            $57$ & 3.6(2) & 3.68 \\
            \hline
        \end{tabular}
    \caption{Polarizability $\mathcal{P}$ of Rydberg S$_{\nicefrac{1}{2}}$ states of $^{40}$Ca$^+$ with the principal quantum numbers 49, 53, and 57. The values obtained from the presented method are compared to the theoretical values from solely numerical calculations in \cite{Kamenski2014}.}
    \end{center}
    \label{tab:results}
\end{table}


To calculate uncertainties of polarizability, we simulated the lineshape properties for thermal and coherent states with respect to their linewidth and their relative shifts as follows. We chose $\left|\alpha\right|$ randomly from a normal distribution with mean and standard deviation calculated according to the experimental values for alpha and the uncertainty from simulations described above (section~\ref{subsec:motional-excitation}). Similarly, temperature values were randomly drawn from a normal distribution around the measured temperature of the ion after state initialization. In addition, the lineshape was calculated with a randomly drawn polarizability in the range of $0.5-2$ times the numerical values in~\cite{Kamenski2014}. The width and relative shifts of these lineshapes were compared for 3000 individual simulations. We estimate the standard deviation of the polarizability as the subset of polarizabilities that yield the same relative shifts and linewidths observed in the experiment. We repeat this procedure for each individual coherent motional excitation and extracted $|\alpha|$. Note that the uncertainty in the polarizability is a function of uncertainty in alpha for a given distribution of temperature. We find that uncertainty for determining the polarizability increases linearly with the magnitude of the polarizability, resulting in a relative uncertainty of $14(2)\%$. By averaging over the resulting values of $\mathcal{P}$ of several spectra, we further increase the accuracy of our measurement.


\section{Conclusion and outlook}
\label{sec:conclusion}

In this work, we have presented a new method for determining polarizability of Rydberg states of trapped ions. We prepared trapped ions in coherent motional states and probed the spectra of subsequently excited Rydberg transitions. We developed a line model for Rydberg lines of a single ion in a Paul trap in which motional degrees of freedom has been taken into account. In our experiments, coherent states of a $^{40}$Ca$^+$ ion with $2<\left|\alpha\right|<12$ and Rydberg transitions for 49, 53 and 57 S$_{\nicefrac{1}{2}}$ states were investigated. Within the experimental accuracy, the extracted polarizabilities using this method are in good agreement with calculated values and follow the expected scaling with the principal quantum number to the power of seven.


The presented results can provide a better understanding of the quadratic Stark effect observed. Tailoring state-dependent trapping frequencies opens up the possibility for investigating structural phase transitions in ion crystals~\cite{Li2012}.
Additionally, microwave dressing of Rydberg S- and P-states with an opposite-sign polarizability can be used for a full control of the polarizability. Such a control will also allow for generating states with even a larger polarizability and and larger transition shifts. This work will also contribute for developing atomic sensors for the electric and microwave fields with high accuracy and sensitivity.

\section{Acknowledgements}
 The authors thank Justas Andrijauskas for his contribution to the lineshape model at an early stage and Dr.~Han Bao for helpful discussions. This work was supported by the Deutsche Forschungsgemeinschaft (DFG) within the SPP 1929 Giant interactions in Rydberg Systems (GiRyd) and the QuantERA grant ERyQSenS. A. M. acknowledges the funding from the European Union's Horizon 2020 research and innovation programme under the Marie Skłodowska-Curie grant agreement No.~796866 (RydIon).

\section*{References}
\bibliographystyle{iopart-num}
\bibliography{LiteraturJV}

\end{document}